\documentstyle[twoside,fleqn,espcrc2]{article}

\def\be{\begin{equation}}
\def\ee{\end{equation}}
\input{psfig}
\newcommand{\AmS}{{\protect\the\textfont2
  A\kern-.1667em\lower.5ex\hbox{M}\kern-.125emS}}

\title{Random matrix model approach to chiral symmetry}

\author{J.J.M. Verbaarschot\address{Dept. of Physics, SUNY at Stony Brook, 
        Stony Brook, NY\,11794}}
       
\begin{document}

\begin{abstract}
We review the application of random matrix theory (RMT) to chiral symmetry 
in QCD. Starting from
the general philosophy of RMT we introduce a chiral random matrix model
with the global symmetries of QCD. Exact results are
obtained for universal properties of the Dirac spectrum: i) finite 
volume corrections to valence quark mass dependence of the chiral condensate, 
and ii) microscopic fluctuations of Dirac spectra. Comparisons with
lattice QCD simulations are made. Most notably, the variance 
of the number of levels in an interval containing $n$ 
levels on average is suppressed by a factor $(\log \,n)/\pi^2 n$.
An extension of the random matrix model
model to nonzero temperatures and chemical potential provides us with a 
schematic model of the chiral phase transition. In particular, this elucidates
the nature of the quenched approximation at nonzero chemical potential.
\end{abstract}

\maketitle

\section{INTRODUCTION}
Random matrix theories have been applied to many areas of physics
ranging from nuclear physics \cite{Wigner} 
to quantum gravity \cite{Ginsparg} and neural networks \cite{Sommers}.
In general, one can divide RMT applications into two different groups.
First, as a description of universal fluctuations of an observable 
expressed in terms of its average value. 
For example, the Hauser-Feshbach formula \cite{Hauser,VWZ}
and universal conductance fluctuations \cite{meso}. 
Second, RMT can be used as a schematic model 
for problems involving disorder. Well-known examples in this category 
are the Anderson Model 
\cite{Anderson}, the
Gross-Witten model \cite{GW} and  models for neural networks \cite{Sommers}.

In this lecture we will review the application of RMT to chiral symmetry
in QCD. Before discussing both types of applications, we introduce
the concept of universality within this context.

As has been argued, in particular by Leutwyler and Smilga \cite{LS}, the mass
dependence of the QCD partition function in the 'mesoscopic' range 
\be
\frac 1{\Lambda} \ll V^{1/4} \ll \frac 1{\sqrt{mV}},
\label{range}
\ee
where $m$ is the quark mass, and $\Lambda$ is a typical hadronic scale,
is given by
\be
Z_{\rm eff}(M, \theta) = \int_{U\in SU(N_f)} dU e^{{\rm Re} \,V\Sigma
\,{\rm Tr}\, M U e^{i\theta/N_f}}.
\label{zeff}
\ee
Here, $M$ is the mass matrix, $\theta$ is the vacuum angle, and $\Sigma= 
|\langle \bar q q \rangle |$.
However, QCD is not the only theory that can be mapped onto this effective
partition function.
In section 3, we will introduce a random matrix
model that can be reduced to (\ref{zeff})
as well. This allows us to formulate universality: different theories with the
same low-energy effective partition function have common properties. 
\section{DIRAC SPECTRUM}
Our main focus will be on the spectrum of the Dirac operator,
\be
i\gamma D \phi_k = \lambda_k \phi_k,
\ee
with spectral density $\rho(\lambda) =\sum \delta(\lambda-\lambda_k)$.
The density of small eigenvalues is related to the chiral 
condensate by means of the Banks-Casher formula \cite{Banks-Casher}
\be
|\langle \bar \psi \psi \rangle| = \frac {\pi\rho(0)}V,
\ee
where it is understood that the thermodynamic limit is taken before
the chiral limit. Since $\{\gamma_5, i\gamma D\} = 0$, the nonzero eigenvalues
occur in pairs $\pm\lambda_k$. The smallest eigenvalue is
of order $\lambda_{\rm min} \approx \pi/\Sigma V$. 
In the QCD partition function,
the Dirac eigenvalues and the quark mass
occur  only 
in the combination $m^2+\lambda_k^2$. Therefore, it is natural to expect 
that universal 
features of the Dirac spectrum can only be found in a region consistent
with (\ref{range}) which,  in  terms of the eigenvalues,
can be expressed  as 
$|\lambda_k| \ll \sqrt{\lambda_{\rm min} \Lambda}\ll\Lambda$.

In order to identify a universal quantity, let us consider 
the mass dependence of the QCD
partition function in the range (\ref{range}), 
\be
\frac {\langle \prod_{k,f}(\lambda_k^2+m_f^2)\rangle}{\langle 
\prod_{k,f}\lambda_k^2\rangle}
= Z_{\rm eff}(M, \theta).
\ee
Here, the average $\langle \cdots \rangle$ is over gauge field configurations
weighted by the QCD action.
By expanding both sides in powers of $m_f$ we find an infinite family of
sum-rules \cite{LS}. The simplest sum rule,
\be
\frac 1{V^2} \sum_{\lambda_k > 0} \frac 1{\lambda_k^2} =\frac {\Sigma^2}{4N_f},
\label{sumrule}
\ee
has been verified for the instanton liquid model of the QCD vacuum
\cite{SVR}. If we write the sum in (\ref{sumrule}) as an integral over the 
spectral density and introduce the microscopic variable $u =\lambda V\Sigma$,
we find
\be
\int_0^\infty \frac 1{V\Sigma} \rho\left( \frac u{V\Sigma} \right) \frac {du}{u^2} =
\frac 1{4N_f}.
\label{seven}
\ee
What enters in (\ref{seven}) 
is the microscopic spectral density,  which in the range
$\lambda \ll \sqrt{\lambda_{\rm min} \Lambda}$ 
can be approximated by its thermodynamic limit
\be
\rho_S(u) = \lim_{V\rightarrow \infty} \frac 1{V\Sigma} 
\rho\left( \frac u{V\Sigma} \right).
\ee
The existence of this limit follows form the spacing of the
eigenvalues near zero virtuality as $\Delta \lambda \sim 1/V$ in the broken 
phase.

Our conjecture
is that $\rho_S(u)$  is a universal function that can be obtained from
a random matrix model with the global symmetries of QCD. 
Note that $\rho_S(u)$ is not fixed by $Z_{\rm eff}(M,\theta)$.
Before defining
this model we discuss the well-known application of RMT to  spectra
of complex systems.
\section{SPECTRAL CORRELATIONS OF COMPLEX SYSTEMS}

RMT has a long history of successes in explaining the statistical
properties  of nuclear spectra \cite{Wigner}.
More recently, spectral correlations have been investigated in
the context of quantum chaos \cite{univers}. The starting point is the
observation that the scale of variations of the average spectral density
and the fluctuations of the spectral density separate. This allows us to
unfold the spectrum. From the original spectrum $\{\lambda_k\}$ we construct
an unfolded spectrum $\{\lambda_k'\}$ with average spectral density equal
to 1. This is achieved by $\int^{\lambda_k'} \bar\rho(\lambda) d\lambda 
=\lambda_k$, where $\bar \rho(\lambda)$ is the average spectral density. 

A variety of statistics has been introduced to analyze the spectral 
correlations of the $unfolded$ spectrum. The best known statistic is
the nearest neighbor spacing distribution $P(S)$. A theoretically
simpler statistic is the distribution of the number of levels, $n_k(n)$, 
in an interval of length $n$. In this lecture, we will consider its
variance denoted by $\Sigma_2(n)$, and the $\Delta_3(n)$
statistic \cite{Dyson} obtained by a smoothening of $\Sigma_2(n)$.

The above statistics can be obtained analytically for 
the invariant random matrix ensembles. 
They  are ensembles of Hermitean matrices with
probability distribution given by $P(H) \sim \exp-{\rm Tr} H^\dagger H$. 
Three different invariant random matrix ensembles can be constructed:
the Gaussian Orthogonal Ensemble (GOE) when $H$ is real, the Gaussian
Unitary Ensemble (GUE) when $H$ is complex and the Gaussian Symplectic
Ensemble (GSE) when $H$ is quaternion real. They are characterized by
the Dyson index $\beta$, which is equal to 1, 2 and 4, respectively.
The relevant random matrix ensemble is determined by the anti-unitary symmetry
of the  Hamiltonian $H$. If there is no anti-unitary symmetry
the spectral correlations are given by the GUE. If the anti-unitary 
symmetry operator $A$ satisfies $A^2 =1$, 
it is possible to find a basis in which
$H$ is real, if $A^2 = -1$, one can construct a basis in which $H$ becomes
quaternion real \cite{Wigner}. Spectral correlations are given by the GOE
and the GSE, respectively.

For simplicity we only give approximate results for the above statistics.
The nearest neighbor spacing distribution is well approximated by the
Wigner surmise given by $P(S) \sim S^\beta\exp(-a_\beta S^2)$, where
$a_\beta$ is a constant. For $n\ge 1$ the number variance is given by
$\Sigma_2(n) \sim (2/\pi^2 \beta) \log n$ and the $\Delta_3$-statistic
$\Delta_3(n)\sim\Sigma_2(n)/2$. The RMT results for these statistics should
be contrasted with the results for uncorrelated eigenvalues. Then,
$P(S) = S$, $\Sigma_2(n) = n$ and $\Delta_3(n) = n/15$. In particular,
we wish to emphasize that long range spectral correlations are
strongly suppressed with respect to uncorrelated eigenvalues.

Spectral statistics of a large variety of systems have been compared to RMT.
We wish to mention, the nuclear data ensemble
\cite{Wigner}, the zeros of Riemann's $\zeta$-function \cite{Odlyzko}, the 
correlations of resonances in electromagnetic resonance cavities \cite{Koch}, 
and the spectrum of the Sinai billiard \cite{Bohigas}. 
The main conclusion is that if the
corresponding classical system is chaotic, the microscopic 
correlations of the spectrum within an irreducible subspace
are given by the RMT with the corresponding anti-unitary 
symmetry.

\section{CHIRAL~$\,$RANDOM~$\,$MATRIX~$\,$MODEL}
In this section we will introduce a random matrix model for the QCD partition
function. Our hope is to construct a model that reproduces both the
microscopic spectral density and the correlations in the bulk of the
spectrum. In the spirit of the invariant RMT we construct a model for the Dirac 
operator with the global symmetries of the QCD partition function as input, but
otherwise  gaussian random matrix elements.  
The chiral random matrix model (chRMM) 
that obeys these conditions is defined by
\cite{SVR,V,VZ}
\be
Z_\nu^\beta = \int DW \prod_{f= 1}^{N_f} \det({\cal D} + m_f)
e^{-\frac{N\Sigma^2 \beta}4 {\rm Tr}W^\dagger W},
\label{zrandom}
\ee
where
\be
{\cal D} = \left (\begin{array}{cc} 0 & iW\\
iW^\dagger & 0 \end{array} \right ),
\label{diracop}
\ee
and $W$ is a rectangular  $n\times m$ matrix with $\nu = |n-m|$ and
$N= n+m$. The matrix elements of $W$ are either real ($\beta = 1$, chiral
Gaussian Orthogonal Ensemble (chGOE)), 
complex
($\beta = 2$, chiral Gaussian Unitary Ensemble (chGUE)), 
or quaternion real ($\beta = 4$, chiral Gaussian Symplectic Ensemble (chGSE)).
A precursor of this model was introduced within the framework
of the instanton liquid partition function \cite{early}.
This model reproduces the following symmetries of the QCD partition 
function:
{\it i)} The $U_A(1)$ symmetry. All nonzero eigenvalues of the random matrix
Dirac operator occur in pairs $\pm \lambda$.
{\it ii)}  The topological structure of the QCD partition function. The matrix
${\cal D}$ has exactly $|\nu|\equiv |n-m|$ zero eigenvalues. This identifies
$\nu$ as the topological sector of the model.
{\it iii)} The flavor symmetry is the same as in QCD. For $\beta = 2$, it is
$SU(N_f) \times SU(N_f)$. For $\beta = 1$, it is $SU(2N_f)$, and for
$\beta = 4$, it is $SU(N_f)$.
{\it iv)} The chiral symmetry is broken spontaneously for two or more flavors
according to the pattern \cite{SmV}  $SU(N_f) \times SU(N_f)/SU(N_f)$,
$SU(2N_f)/Sp(N_f)$ and $SU(N_f)/O(N_f)$ for $\beta = 2$, 1 and 4,
respectively, the same as in QCD \cite{Shifman-three}.
The chiral condensate in this model follows from 
the Banks-Casher relation, 
$
\Sigma = \lim_{N\rightarrow \infty} {\pi \rho(0)}/N.
$
($N$ is interpreted as the (dimensionless) volume of space
time.)
{\it v)} The anti-unitary symmetries. For three and more colors with
fundamental fermions the Dirac operator has no anti-unitary symmetries,
and we choose $\beta = 2$ in (\ref{zrandom}).
For $N_c =2$ and fundamental fermions the Dirac operator satisfies
$
[C\tau_2 K, i\gamma D] = 0,
$
where $C$ is the charge conjugation matrix and $K$ is the complex conjugation
operator.
Because, $(C\tau_2 K)^2 =1$, the matrix elements of the Dirac operator
can always be chosen real, and the corresponding random matrix ensemble
is the chGOE.
For two or more colors with fermions
in
the adjoint representation $i\gamma D$ has the symmetry
$
[CK, i\gamma D] =0,
$
but now $(CK)^2 = -1$, which allows us to rearrange the matrix elements of
the Dirac operator into real quaternions. 
The corresponding random matrix ensemble is the chGSE.

The ensemble of matrices (\ref{diracop}) weighted according to (\ref{zrandom})
is also known as the Laguerre ensemble. Note that its 
spectral correlations in the bulk of the spectrum are given by the
invariant random matrix ensemble with the same value of $\beta$ \cite{laguerre}.
\section{UNIVERSAL PROPERTIES OF THE DIRAC OPERATOR}
\subsection{Spectral correlations of the lattice QCD Dirac operator}
Recently, Kalkreuter \cite{Kalkreuter} 
calculated $all$ eigenvalues of the lattice Dirac
operator both for Kogut-Susskind (KS) fermions and Wilson fermions 
for lattices as large as $12^4$. The accuracy of
the eigenvalues was checked via sum rules 
for the sum of the squares of the eigenvalues
the lattice QCD Dirac operator.
\begin{figure*}
%\vspace*{-0.5cm}
%\rule{5cm}{0.2mm}\hfill\rule{5cm}{0.2mm}
%\vskip 2.5cm
%\rule{5cm}{0.2mm}\hfill\rule{5cm}{0.2mm}
\psfig{figure=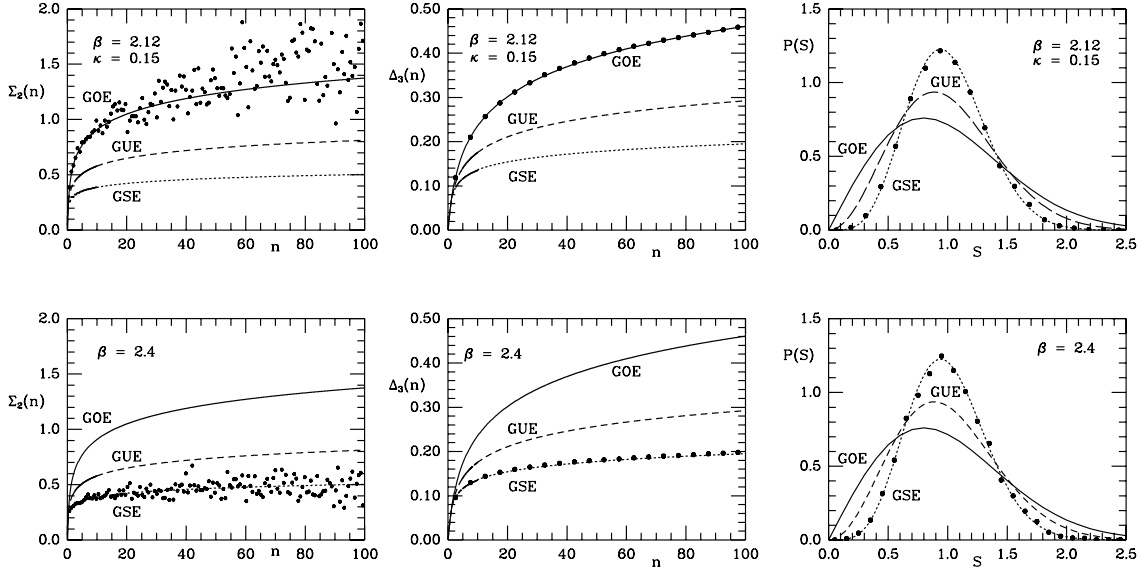,width=150mm,angle=-90}
\vspace*{-0.7cm}
\caption{Spectral correlations of Dirac eigenvalues for Wilson fermions
(upper) and KS-fermions (lower). }
\label{correlations}
\vspace*{-0.5cm}
\end{figure*}
The anti-unitary symmetry of the KS and Wilson Dirac operator is 
different. For KS fermions we have \cite{Teper},
$
[\tau_2 K, D^{KS}] = 0,
$ 
with $(\tau_2 K)^2 = -1$, whereas for the $Hermitean$ Wilson Dirac
operator,
$
[\gamma_5 CK\tau_2, \gamma_5 \gamma D^{\rm Wilson}] = 0,
$ with
$(\gamma_5 CK\tau_2)^2 = 1$. Therefore, we expect that the eigenvalue
correlations are described by the GSE and the GOE, respectively \cite{HV}.
In Fig. \ref{correlations} we show the result for $P(S)$, $\Sigma_2(n)$
and $\Delta_3(n)$. The results for KS fermions are for 4 dynamical flavors
 with $ma = 0.05$ on a $12^4$ lattice. The results for Wilson fermion were
obtained for two dynamical flavors on a $8^3\times 12$ lattice.
The values of $\beta$ and $\kappa$ are given in the label of the figure.
For a discussion of other statistics we refer to \cite{HKV}.

An interesting question is the fate of spectral correlations for
KS fermions in the continuum limit.
To answer this question we have analyzed the
100-200 eigenvalues closest to zero. Even for the
weakest coupling that was studied ($\beta =2.8$)  
no deviation from the GSE was seen.
\subsection{Microscopic spectral density for a liquid of instantons.}

\begin{figure}[ht]
\begin{center}
\leavevmode
%\epsffile{trento34.ps}
\psfig{file=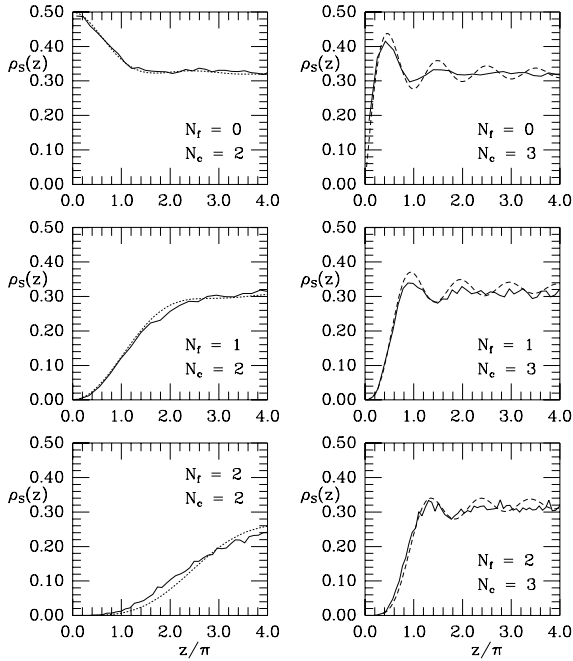,width=75mm}
\end{center}
\oddsidemargin3.7cm
\evensidemargin3.7cm
 \setlength{\textwidth}{16cm}
\vspace*{-1cm}
\caption{The microscopic spectral density for 
a liquid of instantons. }
\oddsidemargin1.7cm
\evensidemargin1.7cm
\setlength{\textwidth}{18cm}
\label{microscopic}
\vspace*{-0.9cm}
\end{figure}

The  investigation of  the microscopic spectral density requires a very
large number of independent gauge field configurations, which is difficult
to generate by a lattice QCD simulations (see however \cite{progress}).
However, it is possible
to simulate a large number of independent 
instanton liquid configurations. If the
microscopic spectral density of such 'smoothened' field configurations
is given by chRMM, one certainly expects that this is the case for 
lattice QCD configurations.  In Fig. \ref{microscopic} we show the spectral
density of the Dirac operator for $N_c = 2$ and 3, and for 0, 1 and 2 massless
flavors \cite{V}. 
In all cases do we find good agreement with the chRMM prediction. 
For $N_c = 3$, $N_f$ flavors and topological charge $\nu$ 
it is given by \cite{V}
\be
\rho_S(u) = \frac u2 \left ( J^2_{a}(u) -
J_{a+1}(u)J_{a-1}(u)\right),
\ee
where $a = N_f + \nu$. The result for $N_c =2$, which is more complicated,
is given in \cite{V2}.  Qualitatively,
the microscopic spectral density has been observed for 
$SU(2)$ staggered fermions \cite{Teper} and 
the lattice Schwinger model \cite{Dilger}.
\subsection{Valence quark mass dependence of the chiral condensate}

\begin{figure}
%\rule{5cm}{0.2mm}\hfill\rule{5cm}{0.2mm}
%\vskip 2.5cm
%\rule{5cm}{0.2mm}\hfill\rule{5cm}{0.2mm}
\psfig{figure=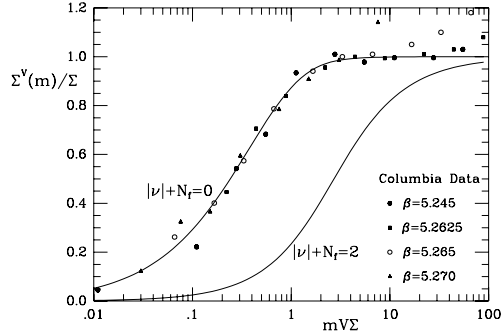,width=65mm,angle=270}
\vspace*{-0.7cm}
\caption{The valence quark mass dependence of the chiral condensate.}
\label{columbia}
\vspace*{-0.9cm}
\end{figure}

An alternative way to probe the Dirac spectrum was introduced by the 
Columbia group \cite{Christ}. They studied the valence quark mass dependence
of the Dirac operator, i.e. $\Sigma(m) = \frac 1N \int d\lambda \rho(\lambda)
2m/(\lambda^2 +m^2)$, for a fixed sea quark mass. In the regime (\ref{range}),
the valence quark mass dependence can be obtained analytically from 
the microscopic spectral density of (\ref{zrandom}) \cite{vplb}
\be
\frac {\Sigma(x)}{\Sigma} = x(I_{a}(x)K_{a}(x)
+I_{a+1}(x)K_{a-1}(x)),
\label{val}
\ee
where $x = mV\Sigma$ is the rescaled mass and $a = N_f+\nu$. 
In Fig. \ref{columbia} we plot this ratio as a function of $x$ for 
lattice data of two dynamical flavors with  mass $ma = 0.01$ and $N_c= 3$ on a
$16^3 \times 4$ lattice.  We observe
that the lattice data for different values of $\beta$ fall on a single curve.
Moreover, in the mesoscopic domain (\ref{range}) 
this curve coincides with the random matrix prediction for $N_f = \nu = 0$.
Apparently, the zero modes are completely mixed with the much larger number of 
nonzero modes. For eigenvalues much smaller than the sea quark mass, we expect
to see the $N_f = 0$ eigenvalue correlations.
\section{SCHEMATIC MODEL THE QCD PARTITION FUNCTION}
In this section we review results for a chiral random matrix model with
nonzero temperature, $T$,  and chemical potential, $\mu$. 
The deterministic parts of this 
model are given by the matrix elements of  $\gamma_0 \partial_0 + \mu \gamma_0$
in a  basis  $\exp[(i((2n+1)\pi -{\rm arg} P)tT)] \phi_n(x)$, where $P$ is the 
Polyakov loop (introduced in \cite{Stephanov1}). If the 
matrix elements of the remaining terms are replaced by the random matrix
(\ref{diracop}) we arrive at the model
\cite{JV,Tilo,Stephanov1,Stephanov2,HJV}
\be
{\cal D} = \left (\begin{array}{cc} 0 & iW +i\Omega_T +\mu \\
iW^\dagger+i\Omega_T + \mu & 0 \end{array} \right ),
\label{diracmatter}
\ee
where $\Omega_T= T \otimes_n((2n+1)\pi -{\rm arg} P){\large\bf 1}$.
Inspired by \cite{Kocic}, 
the simplest model is obtained by keeping
only the lowest Matsubara frequency \cite{JV,Tilo}. 
The random matrix model (\ref{diracmatter}) shows a phase transition in $T$ and
$\mu$. A related RMT as a schematic model of the Wilson Dirac
operator at $T=\mu=0$ was studied in \cite{mac1}.
Alternative
random matrix models for chiral phase
transition have been considered as well. I only mention the study of the
Dirac operator in the Gross-Witten model, where the transition is driven
by the gauge field dynamics \cite{Chandrasekharan}.

\subsection{ ${\bf T\ne 0}$ and ${\bf \mu = 0}$}
The spectral density of the model with only Matsubara frequencies $\pm \pi T$ 
can be obtained analytically from the
solution of a cubic equation \cite{JV,Stephanov1}. It shows
a second order phase transition with critical 
temperature
$T_c = 1/\pi \Sigma$ (in the model with one Matsubara frequency) \cite{JV}. 
Such a transition is typical for a 4-fermion model that is obtained
after integrating over the random matrices. A particular interesting 
feature that  could be checked by recent lattice studies
of the QCD Dirac spectrum \cite{progress,Jansen}, concerns
the fluctuations of the smallest eigenvalue
which is qualitatively different below and above $T_c$ \cite{JV}. 
 
If the phase of the Polyakov line is non-zero,
the smallest Matsubara frequency is lower resulting in a higher
critical temperature \cite{Stephanov1}. 
This result, that can also be obtained from an 
analysis of the NJL model \cite{Meisinger,Huang}, 
explains recent lattice data by the Columbia group \cite{Christ} 
showing that chiral symmetry is restored later for the field 
configurations with a nonzero $Z_{3}$ phase. 

Finally, the model (\ref{diracmatter}) provides additional evidence for the
universality of the microscopic spectral density \cite{zee}. It can be shown 
that \cite{Sener}, 
in spite of the dramatic change of the average spectral density, the
microscopic spectral density does not change below $T_c$.

\subsection{ ${\bf T=0}$ and ${\bf \mu \ne 0}$}

In this section we study the model (\ref{diracmatter}) in the limit 
$T\rightarrow 0$ at finite $N$. Note that the $T\rightarrow 0$ limit 
taken after $N\rightarrow \infty$ is more subtle, $e.g.$,
the Fermi-Dirac distribution is obtained only 
after summing over $all$ Matsubara frequencies \cite{kapusta}.
For nonzero $\mu$ the Dirac operator is no longer Hermitean, 
and its eigenvalues
are scattered in the complex plane. An important question that can be asked
in this context is the nature of the quenched limit. In agreement with
earlier numerical results by Gocksch \cite{Gocksch}, 
Stephanov \cite{Stephanov2} showed analytically for the model 
(\ref{diracmatter}) that the quenched limit is obtained as the limit
$
\lim_{N_f\rightarrow 0} |{\rm det} {\cal D}|^{N_f}.
$
The absolute value can be interpreted as the limit of a model with both
quarks and conjugate quarks which can produce a Goldstone boson with
nonzero baryon number. 
This explains that $ \mu_c \sim \sqrt m$ \cite{everybody}
in quenched lattice QCD simulations
\cite{Stephanov2}. For more details, including results for the spectral
density, we refer to \cite{steph}. A confirmation of some of these results
can be found in \cite{mac2} 
The failure of the quenched approximation at $\mu \ne 0$ was also observed
for a one-dimensional $U(1)$ model
\cite{Gibbs}.

In the unquenched case, the random matrix model shows a phase transition
at $\mu_c = 0.53$ \cite{Stephanov2,HJV}. 
By putting the determinant inside
the operator, the resolvent of the Dirac operator, $G(z) ={\rm Tr} (z-{\cal 
D})^{-1}$, can be obtained numerically. For $z$ inside the domain of the 
eigenvalues, it diverges in the thermodynamic limit, 
and is different from the quenched
result obtained analytically in \cite{Stephanov2}. For $z$ outside this domain
quenched and unquenched results coincide in the thermodynamic limit \cite{HJV}.
The same phenomenon is observed \cite{HJV} for the $U(1)$ model of \cite{Gibbs}.
  
\section{DISCUSSION AND CONCLUSIONS}
We have shown that the spectrum of the QCD Dirac operator shows universal
features that can be obtained from a random matrix model with the global
symmetries of QCD. In this way, we have obtained analytical results for the
finite volume corrections to
the valence quark mass dependence of the chiral condensate and the
spectral correlations in the bulk of the spectrum. 
We have also shown that
an extension of this random matrix model provides a schematic model for
the chiral phase transition. Interesting results have been obtained
for QCD at nonzero temperature and at nonzero chemical potential.

This work was partially supported by the US DOE grant
DE-FG-88ER40388.
\vspace*{-0.2cm}


\begin{thebibliography}{9}
\bibitem{Wigner}
C.E.~Porter, '{\it Statistical theories of spectra: fluctuations}', Academic
Press, 1965; R. Haq, A. Pandey and O. Bohigas,
Phys. Rev. Lett. {\bf 48} (1982) 1086.

\bibitem{Ginsparg}
P. Di Francesco, P. Ginsparg, and J. Zinn-Justin, Phys.\,Rep.\
{\bf 254} (1995) 1.

\bibitem{Sommers}H.J. Sommers, A. Crisanti, H. Sompolinsky and Y. Stein, Phys.
Rev. Lett. {\bf 60} (1988) 1895.
 
\bibitem{Hauser} W. Hauser and H. Feshbach, Phys. Rev. {\bf 87} (1952) 366.

\bibitem{VWZ}J.J.M. Verbaarschot, H.A. Weidenm\"uller and M.R. Zirnbauer,
Phys. Rep. {\bf 129} (1985) 367.

\bibitem{meso} B.L. Altshuler, P.A. Lee and R.A. Webb (eds.), {\it Mesoscopic 
Phenomena in Solids}, North-Holland, New York, 1991; A. Altland and M.R. 
Zirnbauer, cond-mat/9602137.

\bibitem{Anderson}P.W. Anderson, Phys. Rev. {\bf 109} (1958) 1492.

\bibitem{GW}D.J. Gross and E. Witten, Phys. Rev. {\bf D21} (1980) 446.

\bibitem{LS}
H.~Leutwyler and A.~Smilga, Phys. Rev. {\bf D46} (1992) 5607.                  
                          
\bibitem{Banks-Casher}
T.~Banks and A.~Casher,
Nucl. Phys. {\bf B169} (1980) 103.

\bibitem{SVR}E.V. Shuryak and J.J.M. Verbaarschot,
Nucl. Phys. {\bf A560} (1993) 306.

\bibitem{univers}
O.~Bohigas, M.~Giannoni, Lecture notes in Physics
{\bf 209}, Springer Verlag 1984, p. 1;
T.~Seligman, J.~Verbaarschot, and M.~Zirnbauer,
Phys. Rev. Lett. {\bf 53}, 215 (1984);
T.~Seligman and J.~Verbaarschot, Phys. Lett. {\bf 108A} (1985) 183.

\bibitem{Dyson} F.J. Dyson and M.L. Mehta, J. Math. Phys. {\bf 4} (1963) 701.

\bibitem{Odlyzko} A.M. Odlyzko, Math. Comp. {\bf 48} (1987) 273.

\bibitem{Koch}
H.D. Gr\"af {\it et al.}, Phys. Rev. Lett. {\bf 69} (1992) 1296;
S. Deus, P.M. Koch and L. Sirko, Phys. Rev. {\bf E53} (1995) 1146. 

\bibitem{Bohigas}O. Bohigas, M.J. Giannoni and M.J. Schmit, Phys. Rev. Lett.
 {\bf 52} (1984) 1.

\bibitem{V} J. Verbaarschot, Phys. Rev. Lett. {\bf 72} (1994) 2531; Phys. Lett.
{\bf B329} (1994) 351; Nucl. Phys. {\bf B427} (1994) 434.

\bibitem{VZ}J.J.M. Verbaarschot and I. Zahed,
Phys. Rev. Lett. {\bf 70} (1993) 3852.

\bibitem{early}M. Nowak, J.J.M. Verbaarschot and I. Zahed, Phys. Lett. {217B}
(1989) 157; Yu. A. Simonov, Phys. Rev. {\bf D43} (1991) 3534.

\bibitem{SmV}
A. Smilga and J. Verbaarschot, Phys. Rev. {\bf D51} (1995) 829;
M.A. Halasz and J.J.M. Verbaarschot, Phys. Rev. {\bf D52} (1995) 2563.

\bibitem{Shifman-three}
M. Peskin, Nucl. Phys. {\bf B175} (1980) 197;
S. Dimopoulos, Nucl. Phys. {\bf B168} (1980) 69;
M. Vysotskii, Y. Kogan and M. Shifman,
Sov. J. Nucl. Phys. {\bf 42} (1985) 318;
D.I. Diakonov and V.Yu. Petrov, Lecture notes in physics, {\bf 417}, 
Springer 1993.

\bibitem{laguerre}H.S. Leff, J. Math. Phys. {\bf 5} (1964) 763; D. Fox and P.B. 
Khan, Phys. Rev. {\bf 134} (1964) B1152; B.V. Bronk, J. Math. Phys. {\bf 6} 
(1965) 228; T. Nagao and M. Wadati, J. Phys. Soc. Jap. {\bf 60} (1991) 3298,
{\bf 61} (1992) 78, 1910.
 
\bibitem{Kalkreuter}T. Kalkreuter,  Comp. Phys. Comm. {\bf 95} (1996) 1;
Phys. Lett. {\bf B276} (1992) 485; Phys. Rev. {\bf D48} (1993) 1926.

\bibitem{Teper}S.J. Hands and M. Teper, Nucl. Phys. {\bf B347} (1990)
819.

\bibitem{HV}M.A. Halasz and J.J.M. Verbaarschot,
Phys. Rev. Lett. {\bf 74} (1995) 3920.

\bibitem{HKV}M.A. Halasz, T. Kalkreuter and J. Verbaarschot, this proceedings.

\bibitem{progress}R. Wensley, this proceedings; J. Stack and R. Wensley,
in progress; J. Kogut, J.F. Lagae and D. Sinclair, this proceedings. 

\bibitem{V2}J. Verbaarschot, Nucl. Phys. {B426} (1994) 559.

\bibitem{Dilger}H. Dilger, Int. J. Mod. Phys. {\bf C6} (1995) 123.

\bibitem{Christ}S. Chandrasekharan, Lattice 1994, 475;
S. Chandrasekharan and N. Christ, Lattice 1995, 527; N. Christ, this 
proceedings.

\bibitem{vplb}J.J.M. Verbaarschot, Phys. Lett. {\bf B368} (1996) 137.

\bibitem{JV}A.D. Jackson and J.J.M. Verbaarschot, Phys. Rev. {\bf D53} (1996)
7223.

\bibitem{Tilo}T. Wettig, A. Sch\"afer and H.A. Weidenm\"uller,
Phys. Lett. {\bf B367} (1996) 28.

\bibitem{Stephanov1}M. Stephanov, Phys. Lett. {\bf B275} (1996) 249.

\bibitem{Stephanov2}M. Stephanov, Phys. Rev. Lett. {\bf 76} (1996) 4472.

\bibitem{HJV}M.A. Halasz, A. Jackson and J. Verbaarschot, in preparation.

\bibitem{Kocic}A. Kocic and J. Kogut, Nucl. Phys. {\bf B455} (1995) 229.

\bibitem{mac1}J. Jurkiewicz, M. Nowak and I. Zahed {\it et al.}, hep-ph/9603308.

\bibitem{Chandrasekharan} S. Chandrasekharan,  MIT-CTP-2530.

\bibitem{Jansen}K. Jansen and C. Liu, this proceedings.

\bibitem{Meisinger}P.N. Meisinger and M.C. Ogilvie, hep-lat/9512011.

\bibitem{Huang}S. Chandrasekharan and S. Huang, Phys. Rev. {\bf D53} (1996) 
5100.

\bibitem{zee}E. Br\'ezin, S. Hikami and A. Zee,
Nucl. Phys. {\bf B464} (1996) 411. S. Nishigaki, hep-th/9606099.

\bibitem{Sener} A.D. Jackson, M.K. Sener and J. Verbaarschot, Nucl. Phys. 
{\bf B} (1996).


\bibitem{kapusta}J. Kapusta, {\it Finite Temperature Field Theory}, Cambridge
University Press, 1989.

\bibitem{Gocksch} A. Gocksch, Phys. Rev. Lett. {\bf 61} (1988) 2054.

\bibitem{everybody}I. Barbour {\it et al.}, Nucl. Phys. {\bf B275} (1986) 296;
M.P. Lombardo, J. Kogut and D. Sinclair, hep-lat/9511026.

\bibitem{steph}M. Stephanov, this proceedings.

\bibitem{mac2}R. Janik, {\it et al.}, hep-ph/9606329.

\bibitem{Gibbs}P.E. Gibbs, Phys. Lett. {\bf B182} (1986) 369.

 


\end{thebibliography}
\end{document}